\documentclass[onecolumn,superscriptaddress,
nofootinbib,
 amsmath,amssymb,
 aps,
]{revtex4-2}

\usepackage{color}
\usepackage{graphicx}
\usepackage{dcolumn}
\usepackage{bm}
\usepackage{hyperref}
\usepackage{xcolor}

\newcommand{\be}{\begin{eqnarray}}
\newcommand{\ee}{\end{eqnarray}}

\begin{document}

\title{Taming singularities and chaos in conformal gravity}

\author{Jiale Gu}
\affiliation{Center for Astronomy and Astrophysics, Center for Field Theory and Particle Physics, and Department of Physics, Fudan University, Shanghai 200438, China}

\author{Leonardo Modesto}
\affiliation{Dipartimento di Fisica, Universit\'a di Cagliari, Cittadella Universitaria, I-09042 Monserrato, Italy}
\affiliation{INFN, Sezione di Cagliari, Cittadella Universitaria, I-09042 Monserrato, Italy}

\author{Cosimo Bambi}
\email[Corresponding author: ]{bambi@fudan.edu.cn}
\affiliation{Center for Astronomy and Astrophysics, Center for Field Theory and Particle Physics, and Department of Physics, Fudan University, Shanghai 200438, China}
\affiliation{School of Humanities and Natural Sciences, New Uzbekistan University, Tashkent 100001, Uzbekistan}

\date{\today}

\begin{abstract}
We hereby address the cosmological singularity problem in a general gravitational theory invariant under Weyl conformal transformations. In particular, we focus on the Bianchi IX spacetime and we show that both the initial (big bang) and final (big crunch) singularities disappear in an infinite class of conformal frames naturally selected according to analyticity.
It turns out that the past and future singularities are both unattainable within a finite affine parameter (for massless particles) or within a finite proper time (for massive and conformally coupled particles). In order to prove such a statement, we show the geodesic completion of the spacetime when probed by massless, massive, and conformally coupled particles. Finally, the chaotic behavior of the spacetime near the singularity is tamed by a conformal rescaling that turns the Bianchi IX metric into a quasi-FLRW spacetime. 
 \end{abstract}

\maketitle

\section{\label{sec:level1}Introduction}

Since the discovery of the first exact non-trivial solution of Einstein's gravity (EG), namely the Schwarzschild metric, several ideas have been proposed in order to solve the singularity issue. These attempts often involve quantum gravitational effects and/or different theories beyond EG. Indeed, the singularity problem cannot be overlooked because the majority of the solutions are singular in one point or even in extended regions of the spacetime. However, it is exactly this last irrefutable fact that should alert us about the real meaning of singularities. We believe that such a widespread and varied presence of singularities is the sign of a hidden symmetry in the theory. In other words, we should understand EG and its symmetries before quantizing or modifying it.

As a follow up of several previous works~\cite{Bambi:2016wdn,Bambi:2016yne,Bambi:2017yoz,Bambi:2017ott,Chakrabarty:2017ysw,Zhou:2018bxk,Zhang:2018qdk,Zhou:2019hqk,Modesto:2021bup}, we hereby address the singularity issue in theories enjoying Weyl conformal invariance, whose crucial role was pointed out for the first time in Ref.~\cite{narlikar:1977nf} and more recently in Ref.~\cite{tHooft:2009wdx}. In the past, we showed how the Weyl invariance can be implemented in order to make spherically symmetric or axisymmetric spacetimes geodetically complete~\cite{Bambi:2016wdn}. We also solved the big bang singularity problem~\cite{Modesto:2022asj,Calcagni:2022tuz} and, more recently, we discovered a class of stable wormhole solutions sourced by normal not exotic matter~\cite{Cadoni:2025cmf}. Moreover, a proper and unique conformal frame can also explain the galactic rotation curves \cite{Li:2019ksm}\footnote{Hystorically, {\em conformal gravity} refers to the theory proposed by H. Weyl in $1918$ in order to unify  gravity and the Maxwell’s electromagnetic theory~\cite{{Weyl:1918pdp,Weyl:1918ib}}. Such a proposal did not work, but was later resumed as a purely gravitational theory by P.D. Mannheim~\cite{Mannheim:2011ds} and extended by R.K. Nesbet~\cite{Nesbet:2023cvj}. Such a theory is consistent with unitarity, renormalizable, and able to address issues like the galactic rotation curves and the accelerating Hubble expansion. All the results presented in this paper are based on the symmetry and on the Bianchi IX metric as an exact solution, thus they can be exported to any conformal theory sharing the same solution.}.

Contrary to what is stated in most of literature, in order to solve the singularity problem we do not need exotic theories or quantum gravity, but it is sufficient Einstein's conformal gravity, which we will shortly review in Section~\ref{sbsct:3a}. Quantum gravity takes part in the singularity business only in order to preserve the Weyl invariance at quantum level. Indeed, a gravitational theory that is Weyl invariant at classical level and finite at quantum level will be manifestly conformal invariant at quantum level too. Such a theory has a much larger space of solutions at which the conformal symmetry can be spontaneously broken. Hence, since in the conformal phase singularity free solutions are gauge equivalent, requiring analyticity of the solutions will force us to select regular solutions in the broken phase too. Examples of gravitational theories finite at quantum level in the quantum field theory framework were extensively described in Refs.~\cite{Krasnikov:1987yj,kuzmin,Modesto:2011kw, Modesto:2014lga}.

From the technical point of view, the outcome of this paper is shared by any theory invariant under the following Weyl conformal transformation in $D$-dimensions
\begin{eqnarray}
    \hat{g}_{\mu\nu} \rightarrow \Omega^2(x) \hat{g}_{\mu\nu}\, , \quad  \phi \rightarrow \Omega^{\frac{2-D}{2}} \phi \, .
\label{WI}
\end{eqnarray}
In this paper, we focus on the Bianchi IX spacetime and we provide an explicit example of rescaling $\Omega(x)$ that makes singularity-free the curvature invariants and geodetically complete the Bianchi IX metric. As we mentioned above, the presence of a conformal phase ensures the regularity of the spacetime, namely the notion of singularity is not even defined whether the concept of distance is gauge dependent, while the choice of the rescaling has to interpolate analytically between the unbroken and the spontaneously broken phase consistently with geodetic completion.

In addition, we make an attempt to avoid the chaotic behavior in conformal gravity. Indeed, the Bianchi IX metric shows a chaotic behavior near the singularity at $t=0$ \cite{Belinsky:1970ew}, but it is always possible to choose a conformal rescaling that converts the Bianchi IX metric into the FRW spacetime, namely in conformal gravity not isotropic and isotropic spacetime are quasi-equivalent.

Our paper is organized as follows. In Section~\ref{sct:2}, we briefly review the Bianchi IX metric. In Section~\ref{sct:3}, we show how we can solve the singularities of the Bianchi IX spacetime in conformal gravity. In Subsection~\ref{sbsct:3a}, we present the simplest conformal gravity theory in dimension $D=4$. In Subsections~\ref{sbsct:3b} and \ref{sbsct:3c}, we prove the geodesic completeness of the Bianchi IX spacetime under conformal gravity. In Section~\ref{sct:4}, we show how we can cancel the chaos near $t=0$ by choosing a simple conformal factor $\Omega(x)$. Our conclusions are reported in Section~\ref{s-conc}.

\section{Bianchi IX Spacetime}\label{sct:2}

The Bianchi spacetimes describe cosmological models in which the spacetimes are spatially homogeneous but not necessarily isotropic. These models are often used to describe the evolution of the very early Universe. There are nine Bianchi spacetimes (Bianchi~I, II, III, ..., IX). Their classification is based on the 3-dimensional Lie group of isometries that acts simply transitively on the spatial hypersurfaces of homogeneity~\cite{Bianchi:1897,Taub:1950ez}. The Bianchi IX spacetime has the topology of $\mathbb{R} \times \mathbf{S}^3$ and its line element can be written as~\cite{Misner:1969hg, Dechant:2008pb}
\begin{eqnarray}
    ds^2 = -dt^2 + \sum^{3}_{i,j=1} A_{ij}^2 (t) \omega^{i} \otimes \omega^{j} \, .
\label{eqct:1}
\end{eqnarray}
Let us consider its diagonal form, where the matrix $A_{ij}$ is diagonal, $A_{ij} = {\rm diag}(a_{1}, a_{2}, a_{3})$. The 1-forms $\omega^{i}$ are 
\begin{equation}
\begin{aligned}
    & \omega^{1} = dx + \sin ydz \, , \\ &
    \omega^{2} = \cos xdy - \sin x \cos ydz \, , \\ &\omega^{3} = \sin xdy + \cos x \cos ydz \, ,
\end{aligned}
\end{equation}
and the line element can be written as
\begin{equation}
    ds^2 = -dt^2 + \sum^{3}_{i=1} a_{i}^2 (t) \left( \omega^{i} \right)^2 \, .
\label{eqct:1diag}
\end{equation}
In the coordinate system $(t,x,y,z)$, the line element is
\begin{eqnarray}
ds^2 &=& -dt^2 + a_{1}^2 \, dx^2 + \left( a_{2}^{2} \cos^2 x + a_{3}^{2} \sin^2 x \right) dy^2 
+ \left[ a_{1}^{2} \sin^2 y + \left( a_{2}^{2} \sin^2 x + a^{2}_{3} \cos^2 x \right) \cos^2 y \right] dz^2 \nonumber\\
&& + 2 a_{1}^{2} \sin y \, dx dz + 2 \left( a_{3}^{2} - a_{2}^{2} \right) \sin x \cos x \cos y \, dy dz \, .
\label{eqct:5}
\end{eqnarray}

Let us now consider the isometries of the Bianchi IX spacetime. Solving the Killing equation $\mathbf{L}_{\xi} g = 0$, one can derive the following three Killing vectors
\begin{equation}
\begin{aligned}
    & \xi_{1} = \left( 0,\ \sec y \cos z,\  \sin z,\  -\tan y \cos z \right) , \\
     & \xi_{2} = \left( 0,\  - \sec y \sin z,\  \cos z,\  \tan y \sin z \right) , \\ &
      \xi_{3} = \left( 0,\ 0,\  0,\  1 \right) . 
\label{eqct:2}
\end{aligned}
\end{equation}

Instead of the scale factors $a_i(t)$, we can use the Misner variables~\cite{Misner:1969ae, Bojowald:2023fas} to describe the dynamics of the Bianchi IX spacetime
\begin{eqnarray}
    & \beta_{+} = -\frac{1}{2}\ln \left[ \frac{a_{3}}{\left( a_{1} a_{2} a_{3} \right)^{1/3}} \right], \nonumber \\ 
    & \beta_{-} = \frac{1}{2\sqrt{3}} \ln \left( \frac{a_{1}}{a_{2}} \right) .
    \label{Mis1}
\end{eqnarray}
The Misner variables measure the anisotropy of the spacetime. We can introduce the scale expansion $\alpha(t)$ defined as 
\begin{eqnarray}
    \alpha = \frac{1}{3} \ln \left( a_{1} a_{2} a_{3} \right) .
\label{Mis2}
\end{eqnarray}
The scale factors are related the Misner variable as follows
\begin{equation}
\begin{aligned}
    & a_{1}(t) = \exp \left( \alpha(t) +  \beta_{+}(t) + \sqrt{3} \beta_{-}(t) \right) ,\\ 
    & a_{2}(t) = \exp \left( \alpha(t) + \beta_{+}(t) - \sqrt{3} \beta_{-}(t) \right) ,\\ 
    & a_{3}(t) = \exp \left(  \alpha(t) - 2\beta_{+}(t) \right) .
\label{eqct:10}
\end{aligned} .
\end{equation}

If we plug the Bianchi IX metric into the Einstein field equation, we get the following equations 
\begin{eqnarray}
&&
   \frac{a_{1}''}{a_{1}} + \frac{a_{2}''}{a_{2}} + \frac{a_{3}''}{a_{3}} = 8\pi \left( T_{0}^{0} - \frac{1}{2} T \right) \, ,
\label{eqvee0} \\
&&
   \frac{\left( a_{1}' a_{2}a_{3}\right)'}{a_{1}a_{2}a_{3}} - \frac{a_{1}^4 - \left( a_{2}^2 - a_{3}^2\right)^2}{2\left( a_{1}a_{2}a_{3}\right)^2} = 8\pi \left( T_{1}^{1} - \frac{1}{2} T \right) \, ,
\\
&&
   \frac{\left( a_{1}a_{2}'a_{3}\right)'}{a_{1}a_{2}a_{3}} - \frac{a_{2}^4 - \left( a_{3}^2 - a_{1}^2\right)^2}{2\left( a_{1}a_{2}a_{3}\right)^2} = 8\pi \left( T_{2}^{2} - \frac{1}{2} T \right) \, ,
 \\
 && 
 \frac{\left( a_{1} a_{2} a_{3}'\right)'}{a_{1}a_{2}a_{3}} - \frac{a_{3}^4 - \left( a_{1}^2 - a_{2}^2\right)^2}{2\left( a_{1}a_{2}a_{3}\right)^2} = 8\pi \left( T_{3}^{3} - \frac{1}{2} T \right) \, ,
\label{eqvee}
\end{eqnarray}
where the prime $'$ indicates a derivative with respect to $t$ and $T = T_\mu^\mu$ is the trace of the energy-momentum tensor.

\subsection{Dynamics of the Bianchi IX metric}\label{sbsct:2b}

The Bianchi IX metric solves the Einstein field equation $G_{\mu\nu} = 8\pi T_{\mu\nu}$ with a diagonal energy momentum tensor $T^\mu_\nu = {\rm diag} (-\rho,\ p_{1},\ p_{2},\ p_{3})$. In the Misner variables, the components of the Einstein field equation are~\cite{Ryan:1975jw, Lin:1989tv}
 \begin{eqnarray}
  &&   
  \alpha'^2 -  \left( \beta_{+}'^2 + \beta_{-}'^2 \right) - \frac{1}{2} e^{-2\alpha} V = \frac{8\pi}{3} \rho,
\label{eqct:11}
\\
&&
     \alpha'' + \alpha'^2 + 2 \left( \beta_{+}'^2 + \beta_{-}'^2 \right) = -\frac{4\pi}{3} (\rho + p_{1} + p_{2} + p_{3}) ,
 \label{eqct:12}
 \\
 &&
     \beta''_{+} + 3\alpha' \beta'_{+} + \frac{1}{4}e^{-2\alpha} \frac{\partial V}{\partial \beta_{+}} = -\frac{4\pi}{3} \left( 2p_{3} - p_{1} - p_{2} \right) ,
 \label{eqct:122} \\
 &&
     \beta''_{-} + 3\alpha' \beta'_{-} + \frac{1}{4}e^{-2\alpha} \frac{\partial V}{\partial \beta_{-}} = \frac{4\pi \sqrt{3}}{3} \left(p_{1} - p_{2} \right) ,
 \label{eqct:123}
 \end{eqnarray}
where the prime $'$ denotes a derivative with respect to the coordinate $t$ and the potential $V(\beta_{+},\ \beta_{-})$ is defined as
\begin{eqnarray}
    V(\beta_{+},\ \beta_{-}) = \frac{1}{6} \left[ e^{-8\beta_{+}}+2e^{4\beta_{+}}\cosh(4\sqrt{3}\beta_{-}-1)-4e^{-2\beta_{+}}\cosh(2\sqrt{3}\beta_{-}) \right] .
\end{eqnarray}
The two-variables function $V(\beta_{+},\ \beta_{-})$ has the property $V \geq -1/2$ on the plane $\mathbb{R}^{2}_{\beta_{\pm}}$.

In Refs.~\cite{Lin:1989tv, Lin:1990tq}, the authors showed the existence of a final singularity (big crunch) in any Bianchi IX spacetime satisfying the dominant energy condition (DEC). Let us consider Eqs.~(\ref{eqct:11}) and (\ref{eqct:12}). The DEC reads $\rho + p_{1} + p_{2} + p_{3} \geq \rho \geq 0$ and thus Eq.~(\ref{eqct:12}) implies that the second derivative of $\alpha$ is negative, i.e. $\alpha^{\prime \prime} \leqslant 0$. $\alpha^\prime$ must have a maximum value at, say, $t_{\rm max}$ such that for any $t \in [0, +\infty)$, $\alpha' \leq \alpha'(t_{\rm max})$. Hence, the Bianchi IX spacetime cannot expand forever: $\alpha(t)$ must decrease after reaching its maximum and then the universe starts collapsing. This equation shows that there are only two spacetime singularities in the Bianchi IX spacetime: the initial singularity at $t = t_i = 0$ (big bang) and the final singularity at $t=t_{f}$ (big crunch). $t_{f}$ is finite and corresponds to the lifetime of the universe. There are no other singularities between $t_i$ and $t_f$.

Using $\alpha^{\prime}  \leq \alpha'(t_{\rm max})$ in Eq.~(\ref{eqct:11}), we get the following inequality
\begin{eqnarray}
    & \beta_{+}'^2 (t) + \beta_{-}'^2 (t) \leq \alpha'(t_{\rm max})^2  -\frac{1}{2} \left( e^{-2\alpha(t)} V \right)_{\rm max} - \frac{8\pi}{3} \rho _{\rm max} \, .
\label{ieqct:1}
\end{eqnarray} 
We introduce the temporal coordinate $\eta$, defined as
\begin{equation}
    \frac{d\eta}{dt} = e^{-\alpha(t)} \, ,
\end{equation}
and Eq.~(\ref{ieqct:1}) can be rewritten as
\begin{equation}
\begin{aligned}
    \left( \frac{d\beta_{+}}{d\eta} \right)^2 + \left( \frac{d\beta_{-}}{d\eta} \right)^2 & \leq -\frac{1}{2} V + \left(\frac{d\alpha}{d\eta}\right)^2_{max} - \frac{8\pi}{3} \left( e^{2\alpha} \rho \right)_{max} \\ & \leq \frac{1}{4}+ \left(\frac{d\alpha}{d\eta}\right)^2_{max} - \frac{8\pi}{3} \left( e^{2\alpha} \rho \right)_{max} \, .
\end{aligned}
\end{equation}
This inequality shows that the $|\beta_{+}|$ and $|\beta_{-}|$ are all bounded.

\subsection{Chaotic behavior near the singularities}
\label{sbsct:2c}

Previous studies of both the classical \cite{Imponente:2001fy} and the quantum Bianchi IX models \cite{Bojowald:2005cw} showed the presence of chaotic behavior, with a decrease of the chaotic behavior at the quantum level \cite{Bojowald:2023fas}. Although the curvature dominates the overall dynamics near the singularities, the chaotic behavior of the system is extremely sensitive to small perturbations of the energy-momentum tensor. Therefore, for simplicity, it is natural to begin by considering the vacuum Bianchi IX model, which is time-reversal symmetric.

In vacuum, the Einstein field equation (\ref{eqct:11})-(\ref{eqct:123}) is invariant under the coordinate transformation $t \rightarrow t_f - t$ and the solutions of the Misner variables satisfy the following symmetry
\begin{equation}
\alpha(t) = \alpha(t_{f}-t), \quad  \beta_{+}(t) = \beta_{+}(t_{f}-t) , \quad  \beta_{-}(t) = \beta_{-}(t_{f}-t) \, .
\end{equation}
For $t \rightarrow 0$, the evolution of the scale factors is governed by the following two independent equations~\cite{Calcagni:2017sdq}
\begin{eqnarray}\label{eq:a14a}
    & \frac{d^2}{d\tau^2} \ln a_{1} + \frac{1}{2}a_{1}^4 \approx 0 \, ,\nonumber  \\ &
    \frac{d^2}{d\tau^2} \ln a_{2,3} - \frac{1}{2}a_{1}^4 \approx 0 \, ,  
\label{eqct:37}
\end{eqnarray}
where we have introduced the logarithmic time-like coordinate $\tau$, defined by $dt/d\tau = e^{3\alpha} \sim t$. The solution of Eq.~(\ref{eqct:37}) is exactly the Bianchi~I metric (which is also an exact solution of  Einstein field equation in vacuum). The scale factors $a_{1}$, $a_{2}$, and $a_{3}$ for the Bianchi~I spacetime can be expressed as \cite{Belinsky:1970ew}
\begin{eqnarray}
    a_{1} \sim t^{p_{1}},\ a_{2} \sim t^{p_{2}},\ a_{3} \sim t^{p_{3}} \, , 
\label{eqct:7}
\end{eqnarray}
where the exponents $p_{1}$, $p_{2}$, and $p_{3}$ are called the Kasner exponents and can be parametrized in terms of the parameter $u\in(0,1)$, namely
\begin{eqnarray}
    p_{1}(u) = \frac{-u}{1+u+u^2},\quad p_{2}(u) = \frac{1+u}{1+u+u^2},\quad p_{3}(u) = \frac{u(1+u)}{1+u+u^2} \, .
\label{eqct:14}    
\end{eqnarray}
The exponents satisfy the conditions
\be
 && -1/3 < p_{1} <0 < p_{3} < 2/3 < p_{2} < 1 \, , \\
  &&  p_{1}+p_{2}+p_{3}=p_{1}^2+p_{2}^2+p_{3}^2=1 \,  ,
\label{eqct:8}
\\
 &&   p_{1}(u)=p_{1}\left( \frac{1}{u} \right),\quad p_{2}(u)=p_{3}\left( \frac{1}{u} \right),
 \quad p_{3}(u)=p_{2}\left( \frac{1}{u} \right)
\label{eqct:38}
\ee

If now we consider $a_1^4$ as a perturbation, in the limit $\tau \rightarrow -\infty$ near the initial singularity, the solution of Eq.~(\ref{eqct:37}) is
\begin{eqnarray}
    a_{1} \sim t^{|p_{1}|/(1-2|p_{1}|)},\quad  a_{2} \sim t^{(p_{2}-2|p_{1}|)/(1-2|p_{1}|)},\quad a_{3} \sim t^{(p_{3}-2|p_{1}|)/(1-2|p_{1}|)} \, .
\label{eqct:44}
\end{eqnarray}
Defining the three exponents of $t$ in Eq.~(\ref{eqct:44}) as $\widetilde{p}_{1}$, $\widetilde{p}_{2}$, and $\widetilde{p}_{3}$, one can verify that
\begin{eqnarray}
    \widetilde{p}_{1}(u) = p_{2}(u-1),\quad \widetilde{p}_{2}(u) = p_{1}(u-1),\quad \widetilde{p}_{3}(u) = p_{3}(u-1) \, . 
\label{eqct:36}
\end{eqnarray}
From Eq.~(\ref{eqct:36}), we infer that even the exponents $\widetilde{p}_{i}$ satisfy the condition (\ref{eqct:8}) exactly like for the Bianchi~I spacetime, but the positivity of $p_{1}$ and $p_{2}$ is interchanged.

The intervals defined in terms of the parameter $u$ between two sign changes for the exponents $p_{1}$ and $p_{2}$ are called the {\em Kasner epochs}. If we consider the evolution of the spacetime towards the singularity, distances along two axes oscillate and along the third axis decrease monotonically. The intervals between two changes of the axis that decreases monotonically are called the {\em Kasner eras} and each of them contains a certain number of Kasner epochs. These results provide the evolution of scale factors in the Bianchi IX spacetime. The chaotic behavior emerges from the randomness of the number of Kasner epochs in a Kasner era and of the lengths of the Kasner epochs. The preciser properties of this chaotic behavior has been studied before \cite{Belinsky:1970ew, Barrow:1981sx}.

\section{Solving spacetime singularities in conformal gravity}
\label{sct:3}
In this section, we review Einstein's conformal gravity and show how the spacetime singularities in the vacuum Bianchi IX spacetime are tamed by the Weyl invariance.

\subsection{Einstein's conformal gravity}
\label{sbsct:3a}

The $D=4$ Einstein-Hilbert action reads 
\begin{eqnarray}
    S_{EH} = \frac{1}{16\pi}\int d^{4}x\sqrt{-g}  {R} \, , 
    \label{EH}
\end{eqnarray}
and the simplest example of conformal gravity in $D=4$ is obtained by making the replacement 
\be
g_{\mu\nu} \rightarrow 32 \pi \phi^2 g_{\mu\nu}
\label{rep}
\ee
into the action (\ref{EH}). The outcome reads
\begin{eqnarray}
    S = 2 \int d^{4}x\sqrt{-g} \left[ \phi^2 {R} + 6 g^{\mu\nu} \partial_{\mu} \phi \partial_{\nu} \phi \right] \, .
\label{eqct:41}
\end{eqnarray}
The action (\ref{eqct:41}) is invariant under the following rescaling 
\be
    g_{\mu\nu} \rightarrow \Omega^2(x) g_{\mu\nu} \, , \qquad 
    \phi \rightarrow \Omega^{-1}(x) \phi \, .
\label{eqct:25}
\ee
The scalar field $\phi$ (dilaton) is thus necessary to guarantee Weyl conformal invariance.

The equations of motion for theory (\ref{eqct:41}) are obtained by varying the action with respect to the metric and the dilaton field. One obtains
\be
&& \phi^2 {G}_{\mu\nu}  =
   {\nabla}_\nu \partial_\mu \phi^2 - {g}_{\mu\nu} {\Box} \phi^2 
 -  6 \left( \partial_\mu \phi \partial_\nu \phi - \frac{1}{2} {g}_{\mu\nu} g^{\alpha \beta}
 \partial_\alpha \phi \partial_\beta \phi \right) \, ,  \nonumber \\
 &&
 {\Box} \phi = \frac{1}{6} {R} \phi \, . 
 \label{ECGEoM}
 \ee
Einstein's gravity is recovered when the Weyl symmetry is spontaneously broken and the dilaton field assumes the value
\begin{eqnarray}
    \phi^{*} = \frac{1}{\sqrt{32\pi G}} \, .
\label{EfromW}
\end{eqnarray}
In Einstein's gravity, Diff-invariance and, in particular, Lorentz invariance are preserved. The Weyl symmetry is spontaneously broken, but it can be broken to any solution of the equations of motion, not necessarily as in Eq.~(\ref{EfromW}). Note that if $(g_{\mu\nu} , \phi)$ is an exact  solution of Eq.~(\ref{ECGEoM}) even $\left( S(x) g_{\mu\nu} , S^{-1/2}(x) \phi \right)$ is an exact solution of Eq.~(\ref{ECGEoM}). A general solution of the equations of motion breaks spontaneously the Poincare' symmetry, and there is no reason to select the special vacuum in Eq.~(\ref{EfromW}). Therefore, a particular conformal vacuum is spontaneously selected among an infinite number of vacua located on the Weyl's gauge orbit.

Let us now consider the very common set-up of singular solutions in the case of constant vacuum (\ref{EfromW}). In the conformally invariant phase, before breaking the symmetry, we do not have spacetime singularities because singularity free and singular spacetimes are gauge equivalent \cite{Bambi:2016wdn,Bambi:2016yne,Bambi:2017yoz,Bambi:2017ott,Chakrabarty:2017ysw,Zhou:2018bxk,Zhang:2018qdk,Zhou:2019hqk,Modesto:2021bup}. Hence, in order to have an analytic solution that interpolates between the broken and unbroken phases, the Weyl symmetry can only be broken to geodetically complete spacetimes. In the next sections, we will choose the conformal factor $S(x)$ in order to remove the big bang and the big crunch singularities in the Bianchi IX metric. Furthermore, we will explicitly show how conformal symmetry can avoid the chaotic behavior of the Bianchi IX spacetime turning the latter in a quasi-Friedmann-Lemaitre-Robertson-Walker (FLRW) universe. 

Let us here expand on the general singular nature of the rescaling.
In order to remove a coordinate singularity in a diffeomorphism invariant theory, we need a singular general coordinate transformation. The reader can check about the Gullstrand-Painleve’ coordinates' transformation, or similar. 
In exactly the same vain, in a Weyl conformal invariant theory we select a conformal rescaling that is singular in the regions (a point for the Schwarzschild metric, a point or two disconnected points for the FRW spacetime, a ring for the Kerr spacetime, etc.) where the geodesic equation ceases to be valid. Equivalently, geodesics cannot be extended in such regions and the predictability is lost. This procedure mutates an incomplete spacetime into a geodetically complete one making the singularities unreachable. If we choose a conformal rescaling singular in other points or regions, where the metric is perfectly fine, other physical properties can emerge, like for example the splitting of the spacetime into multiple geodetically complete regions~\cite{Chakrabarty:2017ysw,Modesto:2021bup,Modesto:2025cre}. In this respect the causal structure can be drastically affected by a Weyl transformation: casually connected regions can be made causally disconnected by a singular conformal transformation. However, in this paper the geodesic completion of the spacetime is achieved by means of a rescaling that is singular at the point $t=0$ that cannot be reached in finite proper time or finite affine time. Therefore, the causal structure is preserved everywhere besides $t=0$ that becomes the past infinity.

\subsection{Geodesic completion: massless particles} 
\label{sbsct:3b}

The tools introduced in the previous subsection can be directly applied to vacuum spacetimes, so we consider the vacuum Bianchi~IX metric and we choose particular rescaling $S(x)$ to remove its spacetime singularities.

For massless particles, we have
\be
   d \hat{s}^2 = 0  \quad \Longrightarrow \quad \hat{g}_{\mu\nu}\frac{dx^\mu}{d\lambda} \frac{dx^\nu}{d\lambda} = 0 
    \quad \Longrightarrow \quad \hat{g}_{tt} \dot{t}^{2} + \hat{g}_{xx}\dot{x}^{2}+\hat{g}_{yy}\dot{y}^2 + \hat{g}_{zz}\dot{z}^2+2 \hat{g}_{xz}\dot{x}\dot{z}+2 \hat{g}_{yz}\dot{y}\dot{z} = 0 \, , 
\label{eqct:21}
\ee
where the dot $\dot{}$ indicates a derivative with respect to the affine parameter $\lambda$. From the Killing vectors in Eq.~(\ref{eqct:2}), we can construct the following constants of motion
\begin{eqnarray}
    E_{i} = \hat{g}_{\mu\nu} \, \xi_{i}^{\mu} \, \frac{dx^\nu}{d\lambda} \, .
\label{eqct:3}
\end{eqnarray}
Plugging the metric in Eq.~(\ref{eqct:5}) and the Killing vectors in Eq.~(\ref{eqct:2}) into Eq.~(\ref{eqct:3}), we get the following three equations linear in $(\dot{x},\dot{y},\dot{z})$
\be
    && \hspace{-0.5cm}
    E_{1} = \left[\cos y \left(a_{1}^2(t) \sin y \cos z-a_{2}^2(t) \sin x (\sin x \sin y \cos z+\cos x \sin z)+a_{3}^2(t) \cos x (\sin x \sin z-\cos x \sin y \cos z)\right)\right] \dot{z} \nonumber \\ 
    && 
    \hspace{0.5cm} a_{1}^2(t) \cos y \cos z \dot{x} + \left(a_{2}^2(t) 
    \cos x (\sin x \sin y \cos z+\cos x \sin z)+a_{3}^2(t) \sin x (\sin x \sin z-\cos x \sin y \cos z)\right) \dot{y} \, ,  \nonumber \\ 
    && \hspace{-0.5cm}
    E_{2} = \left[\cos y \left(-a_{1}^2(t) \sin y \sin z+a_{2}^2(t) \sin x (\sin x \sin y \sin z-\cos x \cos z)+a_{3}^2(t) \cos x (\cos x \sin y \sin z+\sin x \cos z)\right)\right] \dot{z} \nonumber \\ 
 &&  
 \hspace{0.5cm}  - a_{1}^2(t) \cos y \sin z \dot{x} + \left(a_{2}^2(t) \cos x (\cos x \cos z-\sin x \sin y \sin z)+a_{3}^2(t) \sin x (\cos x \sin y \sin z+\sin x \cos z)\right) \dot{y} \, , \nonumber \\
    && \hspace{-0.5cm}
    E_{3} = 
    \left[a_{1}^2(t) \sin^2 y +\cos^2 y \left(a_{2}^2(t) \sin^2 x+a_{3}^2(t) \cos^2 x\right)\right] \dot{z} +a_{1}^2(t) 
    \sin y \dot{x}+ \left(a_{3}^2(t)-a_{2}^2(t)\right) \sin x \cos x \cos y \dot{y} \, .
\label{eqct:4}
\ee
After solving the system in Eq.~(\ref{eqct:4}) for $(\dot{x},\dot{y},\dot{z})$, we plug the result with the metric in Eq.~(\ref{eqct:5}) into Eq.~(\ref{eqct:21}) to get the following simple equation for $\dot{t}$
\be
 \dot{t}^2 = \frac{f_{1}^2}{a_{1}^2(t)} + \frac{f_{2}^2}{a_{2}^2(t)} + \frac{f_{3}^2}{a_{3}^2(t)} \, , 
\label{eqct:6}
\ee
where $f_{1}$, $f_{2}$, and $f_{3}$ are functions to $x$, $y$, $z$, and $E_i$, namely
\begin{equation}
\begin{aligned}
    & f_{1}(x,y,z;E_{i}) = \cos y \left( E_{1}\cos z - E_{2}\sin z \right) + E_{3}\sin y \, , \\ 
    & f_{2}(x,y,z;E_{i}) = \cos x \left( E_{1}\sin z + E_{2}\cos z \right) - \sin x \left[ \sin y \left( E_{2}\sin z - E_{1}\cos z \right) + E_{3}\cos y \right]  ,  \\ 
    & f_{3}(x,y,z;E_{i}) = \cos x \left[ \sin y \left( E_{2}\sin z - E_{1}\cos z \right) + E_{3}\cos y \right] + \sin x \left( E_{1}\sin z + E_{2}\cos z \right)  .
\label{eqct:34}
\end{aligned}
\end{equation}

Upon integration of Eq.~(\ref{eqct:6}) from the time-like coordinate $t=T$ to $t=0$, we get the total amount of affine time a massless particle needs to reach the singularity
\be
    \Delta \lambda & = & \int_{0}^{T} dt  \left[ \frac{f_{1}^2}{a_{1}^2(t)} + \frac{f_{2}^2}{a_{2}^2(t)} + \frac{f_{3}^2}{a_{3}^2(t)} \right]^{-\frac{1}{2} }\nonumber  \\ 
    & =  & \int_{0}^{T}dt \, e^{\alpha(t)}   \left( f_{1}^2 e^{-2\beta_{+}(t)-2\sqrt{3}\beta_{-}(t)} + f_{2}^2 e^{-2\beta_{+}(t)+2\sqrt{3}\beta_{-}(t)} + f_{3}^2 e^{2\beta_{+}(t)} \right)^{-\frac{1}{2} } \, , 
\label{eqct:13}
\ee
where in the second equality we simply used the Misner's variables (\ref{eqct:10}). It deserves to be notice that the functions in Eq.~(\ref{eqct:34}) depend only on bounded trigonometric functions of the spatial coordinates $(x,y,z)$. Near $t=0$ the integral in Eq.~(\ref{eqct:13}) can be approximated by 
\be
\Delta \lambda & = & 
\int_{0}^{T} dt  \left[ \frac{f_{1}^2}{a_{1}^2(t)} + \frac{f_{2}^2}{a_{2}^2(t)} + \frac{f_{3}^2}{a_{3}^2(t)} \right]^{-\frac{1}{2} }
\simeq 
\int_{0}^{T} dt 
 \left[ \frac{f_{2}^2}{a_{2}^2(t)} + \frac{f_{3}^2}{a_{3}^2(t)} \right]^{-\frac{1}{2} }
 \nonumber \\
 & \simeq &
 \int_{0}^{T} dt 
 \left[  \frac{f_{2}^2}{a_{2}^2(t)} \right]^{-\frac{1}{2} }
= \frac{1}{| f_2 |} \int_{0}^{T} t^{p_2} dt  \propto T^{p_{2}+1}  \,\, \rightarrow \,\, 0   \,\,\, \mbox{for} \,\,\, T \rightarrow 0 \, , 
\label{DeltaLambda}
\ee
where in the first equality we used the conditions $p_1 < 0$ and $p_2, p_3 >0$ while in the second equality we used the condition $p_2>p_3$. The result is that in the standard Bianchi~IX metric a massless particle reaches the singularity for a finite value of the affine parameter $\lambda$ and the spacetime is geodesically incomplete.

\subsubsection{Removing the big bang}
We consider a conformal transformation
\be
\hat{g}_{\mu\nu}^* = S \, \hat{g}_{\mu\nu} 
\ee
and we choose the following rescaling $S$
\begin{eqnarray}
    S = S(t) 
    = 1+ \frac{L^{2n}}{t^{2n}} \, , \quad n > 0 \, .  
\label{eqct:18}
\end{eqnarray}
With such a choice of $S$, we can remove the big bang singularity, as it is shown below. The constants of motion now are
\be
    E_{i}^{*} = S(t) \hat{g}_{\mu\nu} \, \xi_{i}^{\mu} \, \frac{dx^\nu}{d\lambda} = S(t)E_{i}
      \, , \quad  i=1,2,3 \, . 
    \label{Estar}
\ee
We proceed as before: we solve the constants of motion for $(\dot{x}, \dot{y}, \dot{z})$ and we plug the result in $ds^2=0$. We find the counterpart of Eq.~(\ref{eqct:6})
\be
 \dot{t}^2 = \frac{f_{1}^2}{S^2(t) a_{1}^2(t)} + \frac{f_{2}^2}{S^2(t) a_{2}^2(t)} + \frac{f_{3}^2}{S^2(t) a_{3}^2(t)} \, . 
\label{dottRes}
\ee

We want to compute the affine time to reach $t=0$ starting from a point $T$ very close to $t=0$. From the integration of Eq.~(\ref{dottRes}) we get
\be
\Delta \lambda & = & 
\int_{0}^{T} dt  \left[ \frac{f_{1}^2}{S^2(t) a_{1}^2(t)} + \frac{f_{2}^2}{S^2(t)a_{2}^2(t)} + \frac{f_{3}^2}{S^2(t)a_{3}^2(t)} \right]^{-\frac{1}{2} }
\simeq 
\int_{0}^{T} dt 
 \left[ \frac{f_{2}^2}{S^2(t) a_{2}^2(t)} + \frac{f_{3}^2}{S^2(t) a_{3}^2(t)} \right]^{-\frac{1}{2} }
 \nonumber \\
 & \simeq &
 \int_{0}^{T} dt 
 \left[  \frac{f_{2}^2}{S^2(t) a_{2}^2(t)} \right]^{-\frac{1}{2} }
= \frac{1}{| f_2 |} \int_{0}^{T} S(t) t^{p_2} dt  
\simeq \frac{1}{| f_2 |} \int_{0}^{T}  \frac{L^n}{t^{2n}}  t^{p_2} dt
\propto T^{p_{2}- 2n +1} . 
\ee
Now the integral diverges if $p_2 - 2n +1 <0$, which requires $n> \left(p_2+1\right)/2$. Since $p_2 < 1$, we just need $n \geqslant 1$ to have an infinite $\Delta \lambda$. If the massless particle cannot reach the spacetime singularity for a finite value of the affine parameter, we have solved the singularity and the spacetime is now geodetically complete.

\subsubsection{Removing the big crunch} 

Since the Bianchi IX spacetime shows a second singularity at $t_f > 0$ (big crunch), we have to find a proper rescaling of the metric that takes care of both the singular points. A suitable choice of the function $S$ is
\begin{eqnarray}
S(t) = \left( 1 + \frac{L_{1}^{2n}}{t^{2n}} \right)\left[ 1 + \frac{L_{2}^{2n}}{(t_{f} - t)^{2n}}\right] \, , 
\label{eqct:19}
\end{eqnarray}
where $L_{1}$ and $L_{2}$ are two different length scales. In order to have $S \approx 1$ for $L_1 \ll t \ll t_f$, we need that $L_2 \ll t_f$. The rescaling in Eq.~(\ref{eqct:19}) consists in a product of two similar contributions for the big bang and the big crunch singularities because the Bianchi~IX spacetime near $t_f$ is the same as near $t=0$.

Now we have to integrate Eq.~(\ref{dottRes}) from a regular point $t=T'$ to the singularity $t=t_{f}$
\be
\Delta \lambda_{\rm crunch} & = & 
\int_{T'}^{t_{f}} dt  \left[ \frac{f_{1}^2}{S^2(t) a_{1}^2(t)} + \frac{f_{2}^2}{S^2(t)a_{2}^2(t)} + \frac{f_{3}^2}{S^2(t)a_{3}^2(t)} \right]^{-\frac{1}{2} }
\nonumber \\
& = & 
\int_{T'}^{t_{f}} dt  \ S(t) \left[ \frac{f_{1}^2}{ a_{1}^2(t)} + \frac{f_{2}^2}{a_{2}^2(t)} + \frac{f_{3}^2}{a_{3}^2(t)} \right]^{-\frac{1}{2} }
 \nonumber \\
 & \simeq &
 \int_{T'}^{t_{f}} dt  \ \left[ 1 + \frac{L_{2}^{2n}}{(t_{f} - t)^{2n}}\right]  \left[ \frac{f_{1}^2}{ a_{1}^2(t)} + \frac{f_{2}^2}{a_{2}^2(t)} + \frac{f_{3}^2}{a_{3}^2(t)} \right]^{-\frac{1}{2} }
 \nonumber \\
& \simeq & \frac{1}{| f_2 |} \int_{T'}^{t_{f}} \left[ 1 + \frac{L_{2}^{2n}}{(t_{f} - t)^{2n}}\right] (t_{f} - t)^{p_{2}} dt
\nonumber \\
& = & \frac{1}{|f_{2}|} \int_{0}^{t_{f}-T'} \left( 1 + \frac{L_{2}^{2n}}{ t^{2n}}\right) t^{p_{2}} dt
\nonumber \\
& \simeq & \frac{1}{|f_{2}|} (t_{f}-T')^{p_{2}-2n+1} \, ,
\label{crnull}
\ee
where we used the fact that $(1+L_{1}^{2n}/t^{2n}) \rightarrow 1$ near the singularity $t=t_{f}$ and the time-reversal symmetry $t \rightarrow t_{f}-t$. Since the condition to make the integral (\ref{crnull}) divergent is $p_{2}-2n+1 < 0$, we need $n > \left(p_{2}+1\right)/2>1 $ in order to remove the big crunch singularity.

\subsection{Geodesic completion: conformally coupled and massive particles}
\label{sbsct:3c}

In this section, we check the geodesic completion of the spacetime when probed by conformally coupled and massive particles. The action for conformally coupled and massive particles reads
 \cite{Bekenstein:1975ts}:
\begin{eqnarray}
    S_{\rm cp} = -\int \sqrt{- (m + {\rm C} \phi)^2  \hat{g}_{\mu\nu}dx^{\mu}dx^{\nu}} = \int d\lambda \,  \mathcal{L}_{\rm cp} \, , 
\label{eqct:28}
\end{eqnarray}
where $m$ is the particle mass and $C$ is a coupling constant. For $m=0$ and $C \neq 0$, we have the action of a conformally coupled particle. For $m \neq 0$ and $C = 0$, we have the action of a massive particle. Since the action is invariant under the reparametrization $\lambda \rightarrow \lambda^\prime = f(\lambda)$, we can choose the proper time of the particle for $\lambda$. In this case, we have 
\begin{eqnarray}
    \mathcal{L}_{\rm cp} = - \sqrt{ - (m+ {\rm C} \phi)^2 \, \hat{g}_{\mu\nu}\dot{x}^{\mu} \dot{x}^{\nu} } \, , \quad
    \dot{x}^{\mu} = \frac{dx^{\mu}}{d\lambda} \, , \quad \hat{g}_{\mu\nu}\dot{x}^{\mu} \dot{x}^{\nu} = - 1 \, .
\label{eqct:311}
\end{eqnarray}
The 4-momentum of the particle can be directly inferred from the Lagrangian (\ref{eqct:311})
\begin{eqnarray}
    p_{\mu} = \frac{\partial \mathcal{L}_{\rm cp}}{\partial \dot{x}^{\mu}} = \frac{-(m+C\phi)^2 \hat{g}_{\mu\nu}\dot{x}^{\nu}}{\mathcal{L}_{\rm cp}} = \left( m + C\phi \right)\hat{g}_{\mu\nu}\dot{x}^{\nu} \, .
\end{eqnarray}
The constants of motion of the particle are
\begin{equation}
    F_{i} = \xi^{\mu}_{i} p_{\mu} = (m+C\phi)g_{\mu\nu}\xi^{\mu}_{i} \frac{dx^{\nu}}{d\lambda}
    \label{cazzo}
\end{equation}

Using the equation of motion for the particles: 
\be
\quad \hat{g}_{tt} \dot{t}^{2} + \hat{g}_{xx}\dot{x}^{2}+\hat{g}_{yy}\dot{y}^2 + \hat{g}_{zz}\dot{z}^2+2 \hat{g}_{xz}\dot{x}\dot{z}+2 \hat{g}_{yz}\dot{y}\dot{z} = -1 \, . 
\label{figa}
\ee
Solving (\ref{figa}) for $\dot{x}, \dot{y}, \dot{z}$ we get:
\begin{equation}
\begin{aligned}
    \dot{x} & = \cos y \frac{F_{1} \cos z-F_{2} \sin z+F_{3} \tan y}{a_{1}^{2}(t)\left(C \phi + m\right)} \\ & +\cos y\frac{\sin x \tan y (\cos x \sec y (F_{1} \sin z+F_{2} \cos z)-\sin x (-F_{1} \tan y \cos z+F_{2} \tan y \sin z+F_{3}))}{a_{2}^{2}(t) \left(C \phi + m\right)} \\ & - \cos y \frac{ \cos x \tan y (\cos x (-F_{1} \tan y \cos z+F_{2} \tan y \sin z+F_{3})+\sin x \sec y (F_{1} \sin z+F_{2} \cos z))}{a_{3}^2(t)\left( C \phi+m \right)}
\end{aligned}
\end{equation}
\begin{equation}
\begin{aligned}
    \dot{y} & = \frac{ -2\sin x \cos x \sin y \left(a^2_{2}(t)-a^2_{3}(t)\right) (F_{1} \cos z-F_{2} \sin z)+F_{3} \sin (2 x) \cos y \left(a_{2}(t)^2-a_{3}(t)^2\right)}{a_{2}(t)^2 a_{3}(t)^2 \left(C\phi + m\right)} \\ & + \frac{2 (F_{1} \sin z+F_{2} \cos z) \left(a_{2}(t)^2 \sin ^2 x +a_{3}(t)^2 \cos ^2 x\right)}{a^2_{2}(t) a^2_{3}(t) \left(C\phi + m\right)}
\end{aligned}
\end{equation}

\begin{equation}
\begin{aligned}
    \dot{z} & = \frac{ \cos x (\cos x (-F_{1} \tan y \cos z+F_{2} \tan y \sin z+F_{3})+\sin x \sec y (F_{1} \sin z+F_{2} \cos z))}{a^2_{3}(t)\left(C\phi+m\right)} \\ & + \frac{ \sin x (\sin x (-F_{1} \tan y \cos z+F_{2} \tan y \sin z+F_{3})-\cos x \sec y (F_{1} \sin z+F_{2} \cos z))}{a^2_{2}(t) \left(C\phi +m\right)}
\end{aligned}
\end{equation}

Replacing in Eq.~(\ref{figa}), we find the following expression for $\dot{t}$ for the standard (i.e., before rescaling) Bianchi~IX metric
\begin{equation}
\begin{aligned}
    \dot{t}^2 = 1 + \frac{G_{1}^2}{a_{1}^2(t)} + \frac{G_{2}^2}{a_{2}^2(t)} + \frac{G_{3}^2}{a_{3}^2(t)} 
\label{eqct:30}
\end{aligned}
\end{equation}
The functions $G_{j}$ are related to the functions $f_{j}$ in Eq.~(\ref{eqct:34}) by
\begin{eqnarray}
    G_{j} = \frac{1}{m + C\phi} f_{j}(x,y,z;F_{i}) \equiv \frac{\check{f_{j}}}{m+C\phi},\ i,j = 1,2,3 \, .
\end{eqnarray}

We integrate the differential equation in~(\ref{eqct:30}) and we obtain the affine parameter from $t=0$ to $t=T$
\begin{equation}
\begin{aligned}
    \Delta \lambda_{\rm cp} & = \int^{T}_{0} dt\  \left[ 1 + \frac{G_{1}^2}{a_{1}^2(t)} + \frac{G_{2}^2}{a_{2}^2(t)} + \frac{G_{3}^2}{a_{3}^2(t)} \right]^{-1/2} \\ & = \int^{T}_{0}\ dt\ \frac{1} {(m+C\phi)^2} \left[ 1 + \frac{\check{f}_{1}^2}{a_{1}^2(t)} + \frac{\check{f}_{2}^2}{a_{2}^2(t)} + \frac{\check{f}_{3}^2}{a_{3}^2(t)} \right]^{-1/2} \, .
\end{aligned}    
\end{equation}
Since $p_{1}<0<p_{3}<p_{2}$ and assuming $\phi = {\rm const} = 1/\sqrt{32 \pi G}$, we have 
\begin{equation}
\begin{aligned}
    \Delta \lambda_{\rm cp} & = \left( m+ \frac{C}{\sqrt{32\pi G}} \right)^{-2} \int_{0}^{T} dt\ \left[ 1 + \frac{\check{f}_{1}^2}{a_{1}^2(t)} + \frac{\check{f}_{2}^2}{a_{2}^2(t)} + \frac{\check{f}_{3}^2}{a_{3}^2(t)} \right]^{-1/2} \\ & \simeq \left( m+ \frac{C}{\sqrt{32\pi G}} \right)^{-2} \int_{0}^{T} dt\ \left[ \frac{\check{f}^2_{2} a_{1}^2(t) a_{3}^2(t)}{ \left(a_{1}(t) a_{2}(t) a_{3}(t)\right)^2 }\right]^{-\frac{1}{2}} \\ & = \frac{1}{|\check{f}_{2}|}\left( m+ \frac{C}{\sqrt{32\pi G}} \right)^{-2} \int_{0}^{T} t^{p_{2}} dt \propto T^{p_{2}+1} 
\label{cpbf}
\end{aligned}
\end{equation}
and the integral converge since it is at least of the order of $T^{5/3}$ as $T \rightarrow 0$. Thus $t=0$ is a singularity for conformally coupled or massive particles. It is straightforward to check that the same conclusion is true for $t = t_f$.

Let us consider the conformal rescaling in Eq.~(\ref{eqct:19}). The conserved quantities $G_{j}$ become 
 \begin{eqnarray}
    G_{j}^{*} = \frac{1}{m+C/\sqrt{32\pi S}}\frac{\check{f}_{j}}{S}, \,  j=1,2,3
\end{eqnarray}
Eq.~(\ref{eqct:30}) becomes
\be
S(t) \dot{t}_{\rm cp}^2 = 
   1 + S(t) \left( \frac{{G_{1}^{*}}^2}{a_{1}^2(t)} + \frac{{G_{2}^{*}}^2}{a_{2}^2(t)} + \frac{{G_{3}^{*}}^2}{a_{3}^2(t)} \right)  \, .
\label{eqct:32}
\ee
We integrate from $t = 0$ to $t = T > 0$
\begin{equation}
\begin{aligned}
    \Delta \lambda_{\rm cp} & = \int^{T}_{0} dt \left[ \frac{1}{S(t)} +  \left( \frac{{G_{1}^{*}}^2}{a_{1}^2(t)} + \frac{{G_{2}^{*}}^2}{a_{2}^2(t)} + \frac{{G_{3}^{*}}^2}{a_{3}^2(t)} \right) \right]^{-1/2} \\ & = \int_{0}^{T} dt \left[ \frac{1}{S(t)} + \frac{1}{ \left(m+\frac{C}{\sqrt{32\pi G S(t)}}\right)^2}
    \frac{1}{S^2(t)}
    \left( \frac{{\check{f}_{1}}^2}{a_{1}^2(t)} + \frac{{\check{f}_{2}}^2}{a_{2}^2(t)} + \frac{{\check{f}_{3}}^2}{a_{3}^2(t)} \right) \right]^{-1/2} \, .
\end{aligned}
\end{equation}
Near $t=0$ we can write
\begin{equation}
\begin{aligned}
    \Delta \lambda_{\rm cp} & \simeq \int_{0}^{T} dt \left[ \left( 1 + \frac{L_{1}^{2n}}{t^{2n}}\right)^{-1} + \left( 1  +\frac{L_{1}^{2n}}{t^{2n}} \right)^{-2} \sum_{i=1}^{3} \check{f}_{i}^2 a_{i}^{-2}(t) \right]^{-1/2} \\ & \simeq  \int_{0}^{T} dt \left[ \left( 1 + \frac{L_{1}^{2n}}{t^{2n}}\right)^{-1} + \left( 1  +\frac{L_{1}^{2n}}{t^{2n}} \right)^{-2} \sum_{i=1}^{3} \check{f}_{i}^2 t^{-2p_{i}} \right]^{-1/2} \\ & \simeq \int_{0}^{T} dt \left[ \left( 1 + \frac{L_{1}^{2n}}{t^{2n}}\right)^{-1} + \left( 1  +\frac{L_{1}^{2n}}{t^{2n}} \right)^{-2} \check{f}_{2}^2 t^{-2p_{2}} \right]^{-1/2}
\end{aligned}
\end{equation}
If $n > 1$, the integral diverges\footnote{For the first term, we want $-\frac{1}{2}\cdot 2n \leq -1$, so $n \geq 1$. Similarly, for the second term, $-\frac{1}{2}(4n-2p_{2}) \leq -1$, thus $n \geq (1+p_{2})/2>1$. We thus need $n>1$.}.

Following the same strategy, it is straightforward to check that we can also remove the big crunch singularity. In conclusion, we can remove the singularities in the vacuum Bianchi IX metric for massless, massive, and conformally coupled particles by a proper choice of the rescaling $S$. No particle can reach the spacetime singularities within a finite value of the affine parameter (or proper time) and therefore the spacetime is geodesically complete.

\section{Taming chaos}
\label{sct:4}
Now we study how the chaotic behavior near the singularity changes after rescaling. We consider the vacuum solution of Section~\ref{sbsct:2c} and show that, with a proper rescaling of the metric, the universe can be made to expand or shrink in all directions.

For the sake of simplicity, we consider the Bianchi I metric that has the same features as the Bianchi IX metric concerning the chaotic behavior. The Bianchi I line element reads
\begin{equation}
    ds^2 = -dt^2 + t^{2p_{1}}dx^2 + t^{2p_{2}}dy^2 + t^{2p_{3}}dz^2 \, , 
\end{equation}
which is homogeneous but not isotropic. Hence, we consider the following rescaling 
\begin{eqnarray}
    S (t) = 1+ \left( \frac{t}{t_0}\right)^k \, , \quad k \in \mathbb{R} \, ,
\label{eqct:381}
\end{eqnarray}
where $t_0$ is a constant. After the rescaling of the metric, the exponents for each epoch turn into the following triple of numbers
\be
\left( p_{1}(u) + k, \; p_{2}(u) + k,  \; p_{3}(u) + k \right). 
\ee
From the parametrization in Eq.~(\ref{eqct:14}) and the property in Eq.~(\ref{eqct:38}), the maximum gap among the exponents is
\begin{equation}
    \Delta p_{ij} = \!\!\! \max_{\substack{0<u<1 \\ i,j=1,2,3}} \left\vert p_{i}(u)-p_{j}(u) \right\vert = \frac{2\sqrt{3}}{3} \, .
\end{equation}
If $k$ is large and positive, the gap is negligible and the scale factors $a_{i}$ are very close each other. For instance, if we set the exponent $k$ in the definition of $S(t)$ as follows
\begin{equation}
    k(u) = k_{0} + k_{1}u \, , \quad   | k_0 | \gg \Delta p_{ij},
\label{eqct:ch1}
\end{equation}
with $k_{0} < 0$ and $k_{1} \leqslant 0$, then all three exponents can be approximated by $k(u)$ and the rescaled metric expands in all directions whether we approach $t=0$. The three scale factors expand monotonically with approximately the same speed and the spacetime is described by a quasi-isotropic FLRW metric. Indeed, for large $k$ the scale factors are $a_1 \approx a_2 \approx a_3$ and according to Eq.~(\ref{Mis1}) $\beta_+ \approx \beta_- \approx 0$. The line element reads
\be
ds^2 \approx  \left( \frac{t}{t_0}\right)^k \left( -dt^2 + t^{2p_{1}}dx^2 + t^{2p_{2}}dy^2 + t^{2p_{3}}dz^2 \right) 
\approx - d t^\prime + a(t^\prime) \left( dx^2 + dy^2 + dz^2 \right) \, .
\ee
At the same time, since we do not have anymore a sequence of eras, the chaotic behavior is tamed too. The particular rescaling (\ref{eqct:381}) solves even the singularity issue as discussed in Section~(\ref{sct:3}).

\section{Concluding Remarks}\label{s-conc}

In this paper, we showed the geodesic completion of the Bianchi IX metric in conformal gravity. We have two phases in conformal gravity. In the conformal phase, singular and singularity free spacetimes are gauge equivalent, therefore there is no singularity. In the spontaneously broken phase, we have to select a geodetically complete spacetime that interpolates analytically between the two phases. The metric has to be an analytic function consistent with the absence of singularity in the conformal phase. It turns out that of all possible gauge equivalent spacetimes an infinite subclass of them is geodetically complete and share the same physical features. As a particular example, we selected one of the metrics in the conformal gauge orbit and we proved that it is geodetically complete at the big bang as well as at the big crunch, which turn out to be unreachable instants in the past and future of the universe. We studied the propagation of massless, massive, and conformally coupled test-particles and showed the geodesic completion. Physically nothing can reach the infinite past or the infinite future.

We have also shown that we can mitigate the chaotic behavior by the meaning of a Weyl conformal rescaling. Indeed, it is  always possible to turn the Kasner exponents in each epoch to be nearly identical and the universe quasi-isotropic by the meaning of a conformal transformation. Therefore, in the conformal phase the universe has to be isotropic again because a very anisotropic universe and an FLRW universe are quasi gauge equivalent. 

\vspace{0.3cm}

{\bf Acknowledgments --}
This work was supported by the National Natural Science Foundation of China (NSFC), Grant Nos.~W2531002, 12261131497, and 12250610185.

\nocite{*}

\bibliography{aipsamp}

\end{document}